# On the combined gravity gradient components modeling for applied geophysics

Alexey Veryaskin[1] and Wayne McRae

Gravitec Instruments Ltd.

## ABSTRACT

Gravity gradiometry research and development has intensified in recent years to the extent that technologies providing a resolution of about 1 Eötvös per 1 second average shall likely soon be available for multiple critical applications such as natural resources exploration, oil reservoir monitoring and defense establishment. Much of the content of this paper was composed a decade ago, and only minor modifications were required for the conclusions to be just as applicable today. In this paper we demonstrate how gravity gradient data can be modeled, and show some examples of how gravity gradient data can be combined in order to extract valuable information. In particular, this study demonstrates the importance of two gravity gradient components, *Txz* and *Tyz* which, when processed together, can provide more information on subsurface density contrasts than that derived solely from the vertical gravity gradient (*Tzz*).

Keywords: gravity gradient modeling, gravity gradiometry

### 1. Introduction

Gravity gradients can be treated as fields of differential accelerations between two infinitesimally close spatial points. They can be analytically described using a second rank tensor known as the tensor of gravity gradients

$$T_{ij} = \frac{\partial g_i}{\partial x_j} = -\frac{\partial^2 \Phi}{\partial x_i \partial x_j} \tag{1}$$

Where the *i* and *j* subscripts denote one of the orthogonal Cartesian coordinates (*x*, *y*, *z*), $\Phi$ is the scalar gravitational potential, $g_j$ is a component of the gravitational field vector **g** = ($g_x$, $g_y$, $g_z$), and $\partial/\partial x_i$ is the partial derivative with respect to one of the Cartesian coordinates (i.e. $\partial/\partial x$ $\partial/\partial y$, or $\partial/\partial z$). Out of the tensor's nine components, only five among them are totally independent due to their geometrical symmetry ($T_{ij} = T_{ji}$, where $i \neq j$) and the validity of the Laplace equation outside gravitational field sources (Pedersen and Rasmussen, 1990).

For many gravity gradiometry applications, it is the *Tzz* component (i.e. the second vertical derivative of the gravitational potential) of the gravity gradient tensor that is the primary target as the most informative source of information for data interpretation even when other components

---

[1] Correspondence address: School of Physics, University of Western Australia, 35 Stirling Highway, Nedlands, Perth WA6009, Australia; e-mail: vav@cyllene.uwa.edu.au

are measured (Forsberg, 1984; Marson and Klingele, 1993; Bodard *et al.*, 1993). Despite the fact that the FTG technology is readily available (Murphy, 2004), there are a number of gravity gradiometer developments that are aimed at the measurement of only one or a few gravity gradient components followed by the post-processing stage of recalculating the $T_{zz}$ component from the measured ones (While *et al.*, 2006; DiFrancesco, 2007)

In this paper, we shall consider two off-diagonal gravity gradient tensor components, namely the $T_{xz}$ and $T_{yz}$ components, as an alternative to the $T_{zz}$ source of information. The primary constraint on the direct use of $T_{xz}$ and $T_{yz}$ data as a geophysical tool is that the information they contain is complicated and difficult to interpret (Bodard *et al.*, 1993). Nevertheless, by measuring just the two $T_{xz}$ and $T_{yz}$ gravity gradient components, it is possible to get more information about anomalous subsurface density contrasts than that contained in the vertical gravity gradient ($T_{zz}$).

The easiest way to show this is through combined component modeling which makes it possible to retrieve a level of detail equivalent or better than that contained in $T_{zz}$ gradient images (Veryaskin and Fraser, 1996). It is assumed that a gravity gradiometer capable of directly measuring the $T_{xz}$ and $T_{yz}$ gravity gradients is also capable of producing the real time data sets that are discussed in this study. Recently, a novel design of an absolute string gravity gradiometer has been presented which is aimed primarily at measuring simultaneously the $T_{xz}$ and $T_{yz}$ components from a static or mobile platform, and at the possible data processing algorithms described below (Veryaskin, 2000; Veryaskin, 2003).

The aim of this paper is to demonstrate the utility of combined gravity gradient modeling by producing modeled gradient images of equivalent clarity to those found in other studies (Bodard *et al.*, 1993). To achieve this we have chosen comparable density structures to those used in other studies (buried blocks, a salt dome) and two different modeling approaches.

## 2. Combined component modeling

There are a number of ways in which the gradient components $T_{xz}$ and $T_{yz}$ can be combined for modeling. However, for the scope of this paper only two particularly relevant techniques are presented.

The first approach is to use the two gradient components to obtain the third vertical derivative of the gravitational potential (Veryaskin and Fraser, 1996). In this study we have modeled two structures: buried blocks (representing a sharp edge density contrast) and a salt dome (representing an object having a regular cylindrical geometry) with no "edge of model" effects. These two structures were chosen by Bodard *et al.*, (1993), and provide a convenient baseline to compare our calculations.

The second approach is simply to use $T_{xz}$ and $T_{yz}$ gravity gradients as $x$ and $y$ components of a single vector, and hence deriving its absolute value (Veryaskin and Fraser, 1996; Matthews, 2002). This might be valuable for first order interpretation only and so is dealt with in a separate section using only one example: the salt dome model.

## 2.1. The third vertical derivative of the gravitational potential

The third vertical derivative of the gravitational potential can be derived from the Laplace equation outside any material body

$$T_{xx} + T_{yy} + T_{zz} = 0 \tag{2}$$

From Eq.2 one can easily obtain

$$\frac{\partial T_{xx}}{\partial z} + \frac{\partial T_{yy}}{\partial z} = -\frac{\partial T_{zz}}{\partial z} = -T_{zzz} \tag{3}$$

Then, by changing the order of the derivatives, it yields

$$\frac{\partial T_{xz}}{\partial x} + \frac{\partial T_{yz}}{\partial y} = -T_{zzz} \tag{4}$$

By definition, Eq.4 is invariant with respect to angular rotations in the horizontal plane. That means the *Tzzz* has the same value in any Cartesian coordinate system with different orientations of the horizontal *x* and *y* axes.

Assuming that the *Txz* and *Tyz* components can be measured at spatial points

$Ri = \{x_i, y_i\}, i = 1, 2, 3…N$

Eq.4 can be represented in the following discrete form

$$T_{zzz}(i) = -\frac{T_{xz}(i+1) - T_{xz}(i)}{x(i+1) - x(i)} - \frac{T_{yz}(i+1) - T_{yz}(i)}{y(i+1) - y(i)} \tag{5}$$

There are two possible modeling scenarios that we shall consider in this paper. One is two dimensional grid modeling, representing the amount of information contained in the *Tzzz*. For this type of modeling, we make the following assumptions:

Firstly, that the quantities *Txz* and *Tyz* are measured in absolute units in the horizontal *xy*-plane, and that data is plotted at a constant spatial sampling interval $\Delta$ covering a square area of (N+1) × (N+1) points, where N is an integer. In practice this would be difficult to achieve, but data recorded in the field can be interpolated to a spatially regular grid. Secondly, a single reading station $\{i, j\}$ has the coordinates $x_i = \Delta i$, $y_j = \Delta j$ with *i* and *j* running from 0 to N. Finally, as *Txz*(*i*, *j*) and *Tyz*(*i*, *j*) gravity gradients are known or calculated at every point in the grid, by using Eq.5, we can generate a 2-D, N×N array of data, which contains the same amount of information as the *Tzzz* field:

$$GridFieldZZZ[i, j] = T_{xz}(i, j) - T_{xz}(i+1, j) + T_{yz}(i, j) - T_{yz}(i, j+1) \tag{6}$$

In worked examples we found a gradual drop in image clarity as the distance between sampling points is increased. A further contribution to image distortion would be the inability of the survey platform to hold a true course. In that case, we can use the data collected along an arbitrary path with a constant sampling rate $\Delta t$ in the time domain. The quantity that would contain the equivalent amount of information as the $Tzzz$ field, is either Eq.5 or the following one

$$\mathrm{Pr}ofileFieldZZZ[i] = \frac{T_{xz}(i) - T_{xz}(i+1)}{V_x(i) + V_x(i+1)} + \frac{T_{yz}(i) - T_{yz}(i+1)}{V_y(i) + V_y(i+1)} \tag{7}$$

where $t_{i+1} - t_i = \Delta t = $ const, and $\{Vx, Vy\}$ are the platform velocity horizontal components.

The major advantage of using the third vertical derivative of the gravitational potential $Tzzz$ is that it does not contain any non-inertial contributions. When in motion, a vertical gravity gradiometer does measure the following effective $Tzz$ component

$$T_{zz}^{eff} = T_{zz}^{grav} + \Omega_x^2 + \Omega_y^2$$

where $\Omega_x$ and $\Omega_y$ are the platform angular rate horizontal components. As their spatial derivatives by definition are equal to zero, the $Tzzz$ field is free of any motion effects. However, the same applies to Eq.7 only if the sampling interval $\Delta t$ is infinitesimally small. Also, since Eq.6 and Eq.7 contain only differential combinations of $Txz$ and $Tyz$, any instrument systematic errors and zero point drift disappear as well.

It is possible to avoid the platform velocity terms in Eq.7 since it can be brought to the same form as Eq.6 (which we designate below as the *ProfileFieldZZZ*). Assuming the sensitivity axes $X_g$ and $Y_g$ of the gravity gradiometer, which measures $Txz$ and $Tyz$, are both set at the same angle (45 degree) to the instantaneous horizontal projection of the velocity vector of the moving platform (see Fig.5), one can replace Eq.7 with the following one

$$\mathrm{Pr}ofileFieldZZZ[i] = T_{xz}(i) - T_{xz}(i+1) + T_{yz}(i) - T_{yz}(i+1) \tag{8}$$

Consequently, in the case of the profile modeling (single path modeling) Eq.8 is the one we will use in this paper for the combined $Txz$ and $Tyz$ gravity gradient components.

## 2.2. Single vector modeling

A gravity gradiometer, which can measure simultaneously the $Txz$ and $Tyz$ gravity gradient components in their absolute units, can provide a real time output signal as follows

$$T = \sqrt{T_{xz}^2 + T_{yz}^2} \tag{9}$$

The analytical signal always has its single-valued magnitude whether the platform, which carries the gradiometer, changes its orientation in the horizontal plane or not. This means that it is not necessary to know the exact orientation of the local sensitivity $X_g$ and $Y_g$ axes of the gradiometer with respect to any co-ordinates determined, for instance, by a GPS positioning system. The same situation applies only to the vertical gravity gradient $T_{zz}$ among all the other gravity gradient tensor components. Of course, the use of Eq.9 as a real time output would require a near zero inertial environment (a stable table) as $T_{xz}$ and $T_{yz}$ are subjects to angular rate and angular acceleration contributions (Veryaskin, 2000).

## 3. Models

### 3.1. Square grid modeling: buried blocks

A simple 3D buried blocks model was used in Bodard *et al.* (1993). For ease of comparison the same model is used in this study. The main departure from their method of presenting the model is the calculation of the $T_{xz}$ and $T_{yz}$ gradients used in this paper. In Bodard *et al.* (1993) study, each block was made up of a large number of rectangular prisms whose gravitational field is well known in the literature in terms of gravitational acceleration (Nagy, 1966). The total gravitational signature was calculated as the summation of gravitational acceleration of each individual prism over the volume. In this study we simplified the model by choosing a single rectangular block to represent the entire density contrast and by using exact analytical expressions derived for the $T_{xz}$ and $T_{yz}$ gravity gradients above the under laying structure (see Appendix A to this paper).

Apart from this variation, the design of the survey area and the geometry of the model remain the same as Bodard *et al.* (1993). Four square 10 km × 10 km blocks having a density contrast of 1 g/cm$^3$ were set apart diagonally at a distance equal to their diagonal size, and the shallowest block was placed in the left upper corner of the 110 km × 110 km survey plane. The top of each next block was set to the same elevation as the bottom of the previous upper block. As in the original model, the vertical thickness of each block was chosen to be 3 times its top depth, and the shallowest block was chosen to be 100 m thick with a top elevation of 0 m (ground level). The whole structure was then modeled using Eq.6 in a Mathematica script run on a PowerMac computer. A number of density plots were generated assuming different instrument noise specifications and different survey altitudes. No other filtering, image enhancement or integration technique were used on the raw data sets.

Fig.1 shows an image of the buried blocks model in terms of the calculated $T_{zzz}$ field, i.e. the third vertical derivative of the gravitational potential. This is done in order to get an impression of what information is contained in this particular data that can not be measured directly. Again, we used the exact analytical expression derived from the $T_{xz}$ and $T_{yz}$ gradients, obtained in Appendix A to this paper, by analytically calculating their spatial derivatives with respect to the coordinates *x* and *y*, and then by using Eq.4.

In Fig. 2 the same model is presented in terms of the combined $T_{xz}$ and $T_{yz}$ gravity gradients (*GridFieldZZZ* described by Eq.6). The model calculates $T_{xz}(i, j)$ and $T_{yz}(i, j)$ as if measured simultaneously by a noiseless gravity gradiometer at 150 m altitude above the top surface of the shallowest block. The spatial resolution is 1 km, and 110 × 110 data points cover the survey plane.

As Fig.1 and Fig.2 are almost identical, this proves that the information contained in the combined field of the *Txz* and *Tyz* gravity gradients is equivalent to that of the third vertical derivative of the gravitational potential. None of the other gravity gradient tensor components would provide the same information as *Txz* and *Tyz* ones.

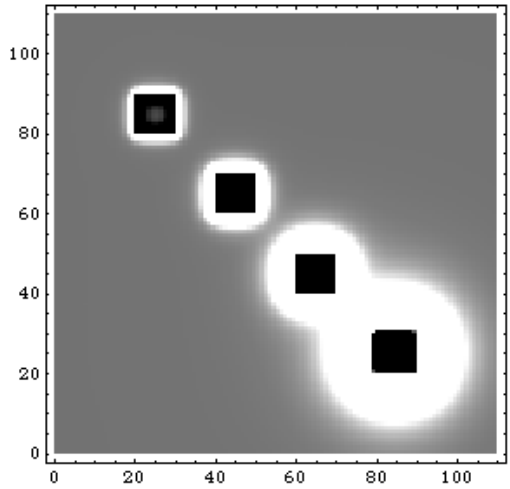

Figure. 1
The density plot of the exact field of the third vertical derivative (*Tzzz*) of the gravitational potential for a set of buried square blocks. Each block has a density contrast of $1$ g/cm$^3$ and measured $10$ km $\times$ $10$ km across in total survey area of $110$ km $\times$ $110$ km. As in Bodard *et al.* (1993) paper, the blocks were placed consecutively deeper in the earth (starting at the surface) at diagonals to each block. The vertical thickness of the shallowest block was 100 m whilst the vertical thickness of subsequent blocks was chosen to be 3 times its top depth. The *Tzzz* field then was evaluated at each kilometer.

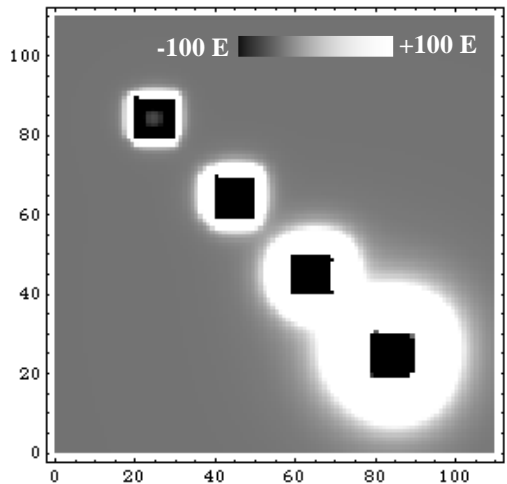

Figure. 2
The same model as in Fig. 1, only presented in terms of the combined *Txz* and *Tyz* gravity gradients (*GridFieldZZZ*($i, j$)) described by Eq.6. It is assumed that *Txz*($i, j$) and *Tyz*($i, j$) are measured simultaneously by a noiseless gravity gradiometer at 150 m altitude above the top surface of the shallowest block. The spatial resolution is 1 km, and $110 \times 110$ reading points cover the survey plane. As Fig.1 and Fig.2 are almost identical, this proves the information contained in the combined field of the *Txz* and *Tyz* gravity gradients is equivalent to that of the third vertical derivative of the gravitational potential.

Here and further in the text the following standard gravity gradient unit is used: $1$ E $= 1$ Eötvös $= 10^{-9}$ s$^{-2}$.

Figure 3.                                   Figure 4.

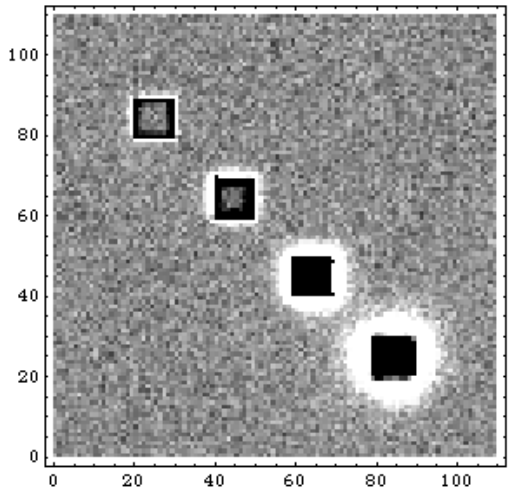 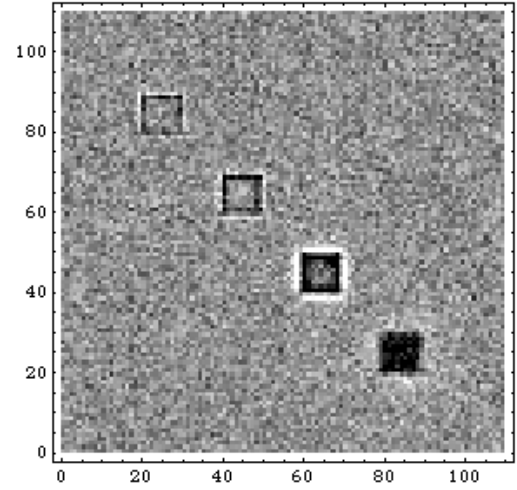

In Fig.3 and Fig.4 noise is added to the data presented in Fig.2, as if the gradiometer measuring *Txz* and *Tyz* gravity gradients was a real device having an operational noise spectral density of N (E/√Hz) in both channels. It is chosen to be 5 E/√Hz in the case of Fig.3 and 50 E/√Hz in the case of Fig.4. As we have said above, the spatial resolution is equal to 1 km. This distance can be covered in about 20 s by flying the gradiometer at about 50 m/s, which approximates an average survey flight speed. That means 20 s average time can be applied to the raw data collected by measuring *Txz* and *Tyz* gravity gradient components. This translates into the signal variance, added to Fig.3, to be about 1 E, and the same in Fig.4 to be about 9 E. Interestingly, a *Txz* and *Tyz* measuring gravity gradiometer with a moderate noise floor of about 50 E/√Hz is capable of detecting the hidden structures corresponding to the buried blocks from an altitude of 150 m.

There are no gravity gradiometers yet capable of achieving 1 E resolution by measuring *Txz* and *Tyz* gravity gradients while on board an aircraft, an airship or a helicopter. However, a resolution of less than 9 E is achievable onboard a low speed (low turbulence) mobile carrier (DiFrancesco, 2007).

### 3.2. Profile modeling: buried blocks

In this section we shall use the Eq.8 instead of Eq.6, which was used in the previous section for the grid modeling. It is assumed that a gravity gradiometer having its local sensitivities axes ($X_g$ and $Y_g$) constantly kept at 45-degree angle to the instantaneous horizontal projection of the velocity vector of the carrier platform. Measurements of *Txz* and *Tyz* gravity gradient components are taken at a constant sampling interval $\Delta t$ in the time domain along a single path as depicted in Fig.5.

In a worked example, however, we shall use the buried blocks model and the data collected through a single profile as shown in Fig.6. The profile crosses the second shallowest buried block along its central horizontal line, i.e. – at 65 km from the origin of the survey plane. The *ProfileFieldZZZ*(*i*) data then was calculated using Eq.8. The result is shown in Fig.7.

Figure 5.      Figure 6.

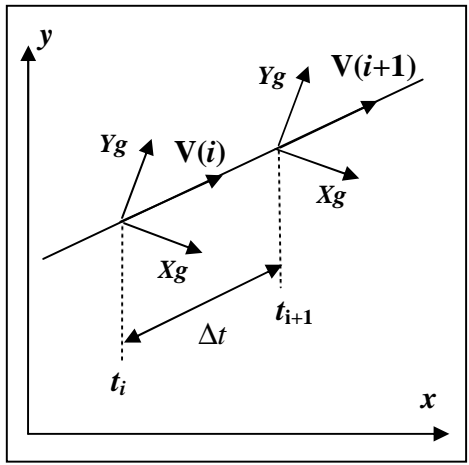
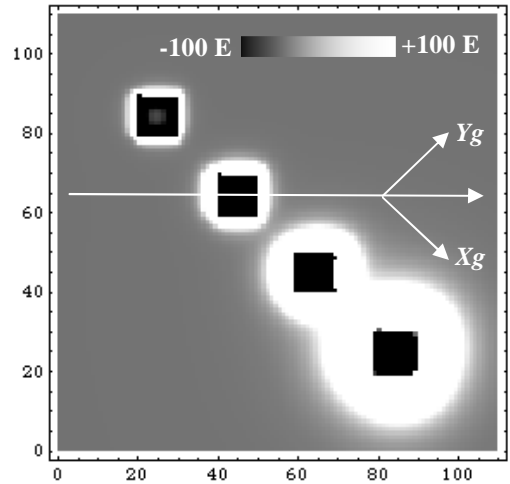

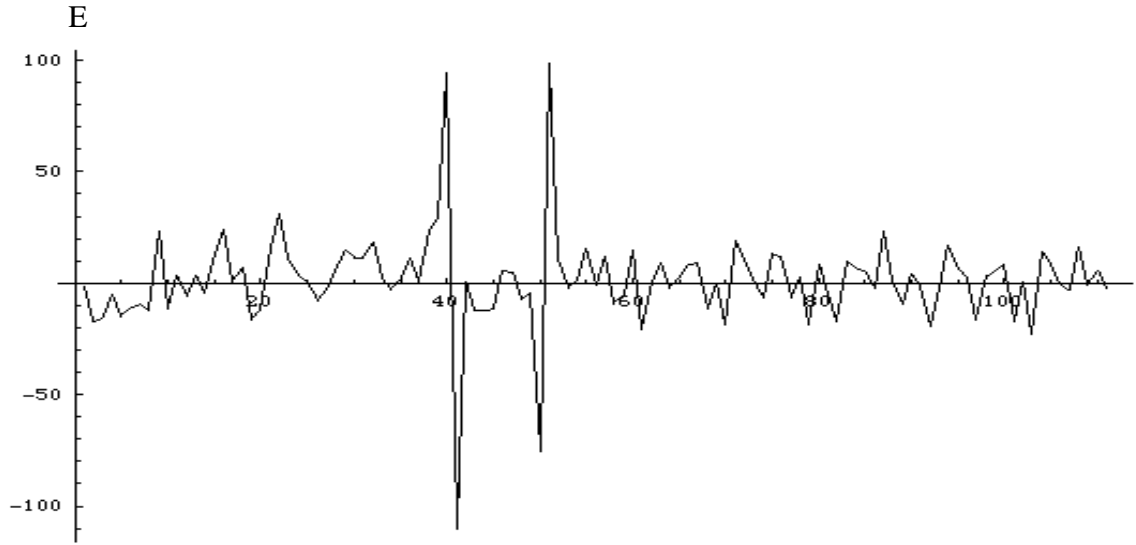

Figure 7. The *ProfileFieldZZZ* data is assumed to be collected at a shortest possible sampling period in the time domain at 50 m/sec survey speed along the straight line shown in Fig.6. The survey line crosses the second shallowest buried block of the model under consideration. Again, 20 s average then is applied to the data, which corresponds to 1 km spatial resolution.

As depicted in Fig.7, despite the moderate sensitivity of the gradiometer that is assumed to be measuring the data (50 E/√Hz spectral noise density is considered in each channel), there is a distinct signal with sharp changes just over the boundaries of the buried block target.

It is important to point out again that the shorter the sampling interval, the less the influence of platform motion on the combined *Txz* and *Tyz* data described by Eq.8. As the information contained in the *ProfileFieldZZZ* is equivalent to that of the exact static *Tzzz* field free of any motion effects, the same should apply to Eq.8.

### 3.3. Square grid and profile modeling: salt dome

Salt domes represent geological structures of special interest for explorers, as they can create structural positions and seals, which capture and contain hydrocarbons, which migrate from oil-bearing rock (Nettleton, 1976). They also act as a good structure for modeling as they are often circular in shape when viewed from above.

Bodard *et al.*, (1993) used a salt dome model to map gravity gradient tensor components. In their study a square $10 \text{ km} \times 10 \text{ km}$ area was covered by $126 \times 126$ points where different gravity gradients were evaluated and the results were then presented in the form of density plots. The salt dome was placed in the centre of the area with its top situated at 1 km below the ground level. As in their previous model, the salt dome was built up from a large number of 0.5 km cubes. Gravity gradients were then evaluated by summing all gravitational signatures produced by the cubes. However, an "edge of model" effect was found as a consequence that cubes could not accurately represent the regular round plan of the model. As a result of that there might be excess values of gravity gradients from sharp boundaries, such as the edges of cubes.

In order to avoid this we used a modified salt dome model comprised of cylindrical disks (see Fig.8) having different thicknesses. Once again, exact analytical expressions for the $T_{xz}$ and $T_{yz}$ gravity gradients in the case of an arbitrary cylindrical gravitational field source were used to represent a single disk (see Appendix B to this paper). The total gravitational signature in Eq.5 was evaluated by a simple summation over all disks. The modified model was chosen to have its top situated at 1.5 km below the ground at the centre of a $10 \text{ km} \times 10 \text{ km}$ square area and the bottom of the lowest disk was situated at 8.5 km below the top of the model. An arbitrary vertical cross section profile of the salt dome model can be generated by using a simple function incorporating its largest diameter and the depth of its top and bottom. A table of disk thicknesses was built up using a condition that each upper disk has its radius set 25 m smaller than that of the next deeper disk, and all these radii fit the profile chosen. A total of 100 disks were found to represent the modified model, which was assigned to have a uniform density contrast of – 0.2 g/cm and the largest diameter of 5 km (Nettleton, 1976).

This approach also allows for individual density contrast values to be assigned to each disk enabling finer details. The result of this modeling is presented in Fig.8, which is comparable with any vertical gravity gradient ($T_{zz}$) images that can be generated under the same conditions (Bodard *et al.*, 1993).

### 3.4. Single vector modeling: salt dome

The salt dome model discussed above can also be used in order to generate a density plot using the analytical signal described by Eq.9. The centre of the salt dome was placed again in the centre ($x = 5$ km, $y = 5$ km) of a $10 \text{ km} \times 10 \text{ km}$ square area. The analytical signal was then evaluated at every 100 m from an altitude of 2000 m. The signal's behavior across the centre of the anomalous area along the *x*-axis is also presented in Fig.10.

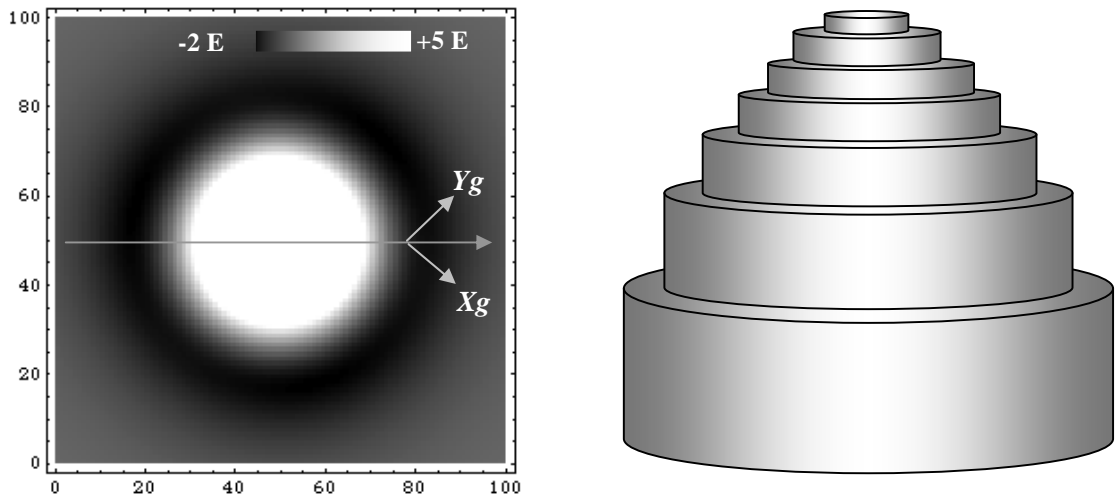

Figure 8. A noiseless density plot (*GridFieldZZZ* described by Eq.6) and the composition of the salt dome model described above. The data was evaluated at every 100 m of the survey 10 km × 10 km area, and the survey plane altitude is 150 m.

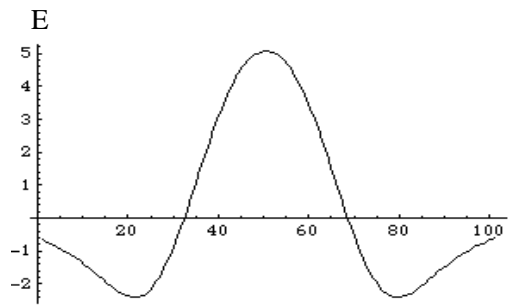

Figure 9.
The same model as in Fig.8 only presented in terms of the profile signal described by Eq.8. The survey line crosses the centre of the salt dome.

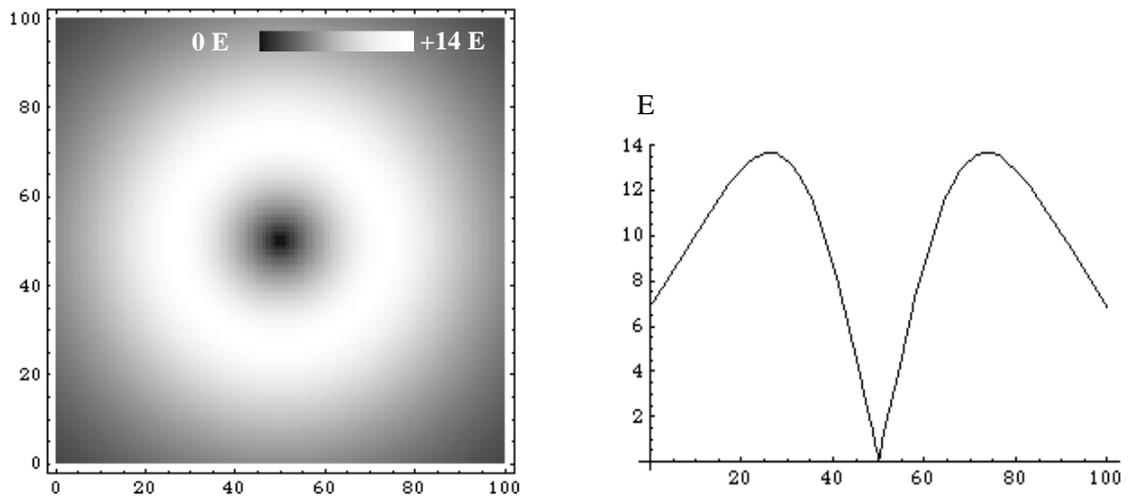

Figure 10. A noiseless density plot of the same model as in Fig.8 only presented as the analytical signal described by Eq.9. A cross-sectional behavior of the signal is also shown on the right-side picture above. The survey plane altitude is 2000 m.

It can be seen that the gravitational signature has a deep minimum just at the centre of the model. One can prove that both the density plot image and the vertical cross section behavior are typical for the gravitational signal described by Eq.9, provided that the gravitational source is a single symmetrical 3-D body, or a cluster of similar structures, which are detected from a high altitude. When the horizontal dimensions of an arbitrary body (or of a cluster of such bodies) becomes of the same order of magnitude as the altitude of the survey plane, then the gravitational signature of the shape is distorted by the nonlinearity of the analytical signal. Non-symmetrical shapes will have more disrupted images but their gravitational signatures will still have minima that can be used in first order interpretation.

## 4. Discussions

This paper clearly demonstrates the amount of information contained in the two $Txz$ and $Tyz$ gravity gradient components. On their own, neither of these two gradient fields looks very informative, when compared with the field of the vertical gravity gradient $Tzz$, as shown by Bodard *et al.* (1993). However it should be evident that the combined component modeling with these two gradients can show at least an equivalent or even better level of information as the single $Tzz$ field does.

It also should be pointed out that a recalculation of any derivatives of the gravitational potential into the fields of their higher-order derivatives, or into some other derivatives of the same order cannot add any new information. Moreover, like any passive filter, this procedure decreases the total amount of information but redistributes it in a way that might be helpful for interpretation (While *et al.*, 2006).

Nevertheless, the potential benefit of using $Tzzz$ data derived from low order derivatives has already been demonstrated (Sagitov *et al.*, 1983). In their study a set of first order Bouguer gravity anomalies over a deepwater subsalt prospect in the Gulf of Mexico were used to calculate the field of the third vertical derivative of the gravitational potential. The $Tzzz$ field turned out to be complicated: instead of one maximum there appeared to be several. These maxima did however coincide with the location of oil deposits, whilst three wells drilled near the maximum of the Bouguer anomalies were found empty.

In this study however, we do not recalculate the $Tzzz$ field from any low rank derivatives. We do combine the individually measured $Txz$ and $Tyz$ gradients in a way that radically redistributes the information contained in each of the individual components, and that information is effectively equal to that of contained in the $Tzzz$ field. The main problem of combined component modeling is that noise increases when the individually measured gravity gradient components are combined.

The other modeling procedure described in this paper (single vector modeling) is less dependent on the individual gradient components being measured with sufficient accuracy. The main reason for this is that at high altitude an instrument is simply being used as a "likely target" detector and is not required to map fine details. The other natural advantage of high altitude flying is the low air turbulence, which allows smoother and faster survey flights.

So, one of the possible ways of using combined components data sets would therefore be a combination of the high-altitude "single vector" minima search, for large scale gravity reconnaissance, followed up by a more accurate low altitude *FieldZZZ* survey over an area of interest.

**Acknowledgements**

We would like to thank Dr Ed Biegert for his comments on the subject of this paper made long time ago during Gravitec's first presentation at a conference of the SEG (Denver, 1996). We also would like to thank Dr Neil Fraser for his contribution at the time the paper was composed as a preprint. Our special acknowledgement to Mr Howard Golden for his comments and improvements he has made in the latest edition of the paper.

**References**

Bodard, J.M., Creer, J.G. and Asten, M.W., 1993. Next Generation High Resolution Airborne Gravity Reconnaissance in Oil Field Exploration. Energy Explor. Exploit., 11(3/4), 198-234.

DiFrancesco, D., 2007. Advances and Challenges in the Development and Deployment of Gravity Gradiometer Systems. EGM International Workshop: Innovation in EM, Grav and Mag Methods. Capri, Italy.

Dransfield, M.H., 1994. Airborne Gravity Gradiometry. PhD thesis, University of Western Australia.

Forsberg, R., 1984. A study of terrain reductions, density anomalies and geophysical inversion methods in gravity field modeling. Report No.355 of the Department of Geodetic Science and Surveying. Ohio State University, 129 p.

Klingele, E.E., Marson, I. and Kahke, H.G., 1993. Automatic interpretation of gravity gradiometric data in two dimensions: Vertical gradient. Geophys. Prosp., 39, 407-434.

Marson, I. and Kingele, E.E., 1993. Advantages of using the vertical gradient of gravity for 3-D interpretation. Geophysics, 58, 1588-1595.

Matthews, R., 2002. Mobile Gravity Gradiometry. PhD Thesis, University of Western Australia.

Murphy, C.A., 2004. the Air-FTG$^{TM}$ airborne gravity gradiometer system. Abstracts from the ASEG-PESA Airborne Gravity 2004 Workshop, 7-11.

Lockerbie, N.A., Veryaskin, A.V. and Xu, X., 1993. Differential gravitational coupling between cylindrical-symmetric, concentric test masses and an arbitrary gravitational source: relevance to the STEP experiment. Classical and Quantum Gravity, 10, 2419-2430.

Nagy, D., 1996. The gravitational attraction of a right rectangular prism. Geophysics, 31, 362-371.

Nettleton, L.L., 1976. Gravity and magnetics in oil prospecting. McGraw-Hill, Inc., New York.

Pedersen, L.B., Rasmussen, T.M., 1990. The gradient tensor of potential field anomalies: some implications on data collection and data processing of maps. Geophysics, 55, 1558-1556.

Prudnikov, A.P., Brychkov, Yu.A. and Marichev, O.I., 1986. Integrals and Series, vols 1 and 2, Oxford: Gordon and Breach.

Sagitov, M.U., Nazarenko, V.S. and Veryaskin, A.V., 1983. Principles and Methods of gravitational field study. Geodeziya i Kartographiya, 11, 11-19.

Veryaskin, A. and Fraser, N., 1996. On the combined gravity gradient components modeling for applied geophysics. Preprint G4-96, Gravitec Instruments, Auckland, New Zealand.

Veryaskin, A., 2000. A novel combined gravity and magnetic gradiometer system for mobile applications. Extended abstracts: Conference of the SEG, Calgary 2000, pp 420.

Veryaskin, A., 2003. String Gravity Gradiometer: Noise, Error Analysis and Applications. Abstracts: AGU+EGS Conference, Nice, France.

While, J., Jackson, A., Smit, D. and Biegert, E., 2006. Spectral analysis of gravity gradiometry profiles. Geophysics, 71, 11-22.

**Appendix A: *Txz* and *Tyz* gravity gradients above an arbitrary rectangular block**

One can show (Lockerbie *et al.*, 1993) that the gravitational potential above a rectangular block of mass density $\sigma$, lateral dimensions $2a \times 2b$ and thickness $d$ (see Fig.1a) can be represented as follows

$$\Phi(\mathbf{R}) = -G\sigma \int_0^\infty dk\, e^{-kz} \left( \int_{-a}^{a} dx' \int_{-b}^{b} dy'\, J_0(k|\mathbf{r}-\mathbf{r}'|) \right) \left( \int_{-H-d}^{-H} dz'\, e^{kz'} \right)$$
$$= -G\sigma \int_0^\infty \frac{dk}{k} \left( \int_{-a}^{a} dx' \int_{-b}^{b} dy'\, J_0(k|\mathbf{r}-\mathbf{r}'|) \right) \left( e^{-k(z+H)} - e^{-k(z+H+d)} \right) \quad (1A)$$

Hereafter $G$ is the gravitational constant, $J_n(x)$ is the ordinary Bessel function of the order n, and **r** is a polar 2-D radius-vector of the spatial point $\{x, y, 0\}$:

$$|\mathbf{r}-\mathbf{r}'| = \sqrt{(x-x')^2 + (y-y')^2}$$

The primed integration is carried out over the volume of the block with its top situated at the depth $H$ below the horizontal *xy*-plane, and the vertical *z*-axis is chosen to go through the centre of mass of the block.

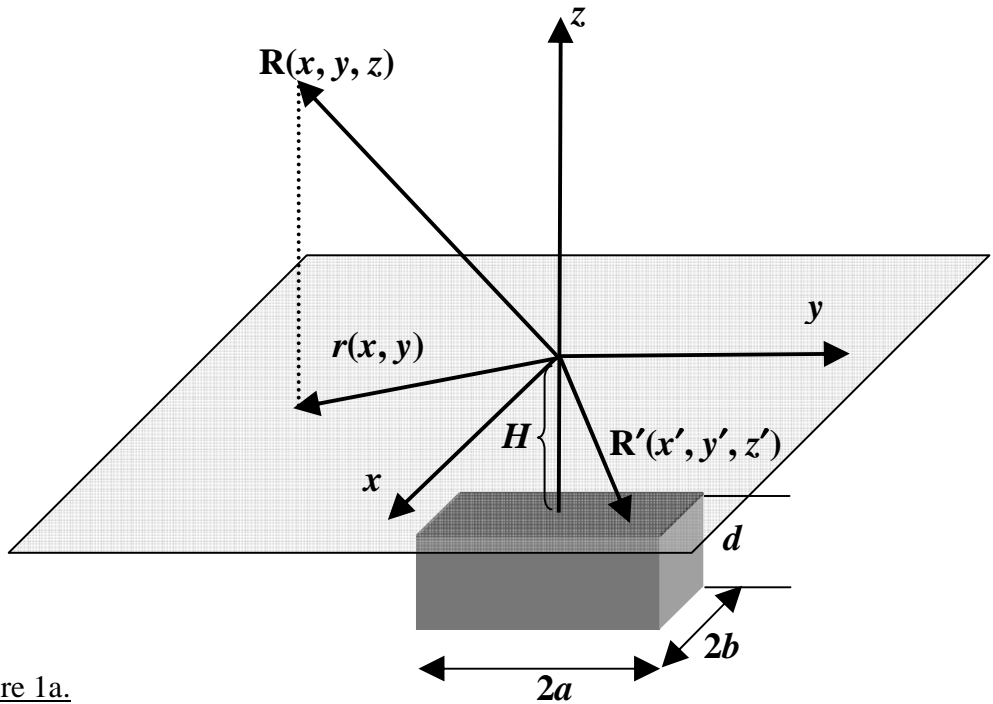

Figure 1a.

By differentiating Eq.2A with respect to the coordinates $z$ and $x$, one can obtain

$$\begin{aligned} T_{xz} &= -G\sigma \int_0^\infty dk \left( \int_{-b}^{b} dy' \int_{-a}^{a} dx' \frac{\partial}{\partial x} J_0(k\sqrt{(x-x')^2 + (y-y')^2}) \right) \left( e^{-k(z+H)} - e^{-k(z+H+d)} \right) \\ &= G\sigma \int_0^\infty dk \left( \int_{-b}^{b} dy' \int_{-a}^{a} dx' \frac{\partial}{\partial x'} J_0(k\sqrt{(x-x')^2 + (y-y')^2}) \right) \left( e^{-k(z+H)} - e^{-k(z+H+d)} \right) \\ &= G\sigma \int_{y-b}^{y+b} dp \left( \frac{1}{\sqrt{p^2 + (x-a)^2 + (z+H)^2}} + \frac{1}{\sqrt{p^2 + (x+a)^2 + (z+H+d)^2}} \right. \\ &\quad \left. - \frac{1}{\sqrt{p^2 + (x+a)^2 + (z+H)^2}} - \frac{1}{\sqrt{p^2 + (x-a)^2 + (z+H+d)^2}} \right) \end{aligned}$$

(2A)

The last integration in Eq.2A yields

$$\begin{aligned} T_{xz} = G\sigma & \left( \ln \frac{\sqrt{(x-a)^2 + (y+b)^2 + (z+H)^2} + y+b}{\sqrt{(x-a)^2 + (y-b)^2 + (z+H)^2} + y-b} \right. \\ & + \ln \frac{\sqrt{(x+a)^2 + (y+b)^2 + (z+H+d)^2} + y+b}{\sqrt{(x+a)^2 + (y-b)^2 + (z+H+d)^2} + y-b} \\ & - \ln \frac{\sqrt{(x+a)^2 + (y+b)^2 + (z+H)^2} + y+b}{\sqrt{(x+a)^2 + (y-b)^2 + (z+H)^2} + y-b} \\ & \left. - \ln \frac{\sqrt{(x-a)^2 + (y+b)^2 + (z+H+d)^2} + y+b}{\sqrt{(x-a)^2 + (y-b)^2 + (z+H+d)^2} + y-b} \right) \end{aligned}$$

(3A)

where the following tabulated integrals were used (Prudnikov *et al.*, 1986)

$$\int_0^\infty dk J_0(kA) e^{-kB} = \frac{1}{\sqrt{A^2 + B^2}} , \qquad \int_{y-b}^{y+b} \frac{dp}{\sqrt{p^2 + \xi^2}} = \ln \frac{\sqrt{(y+b)^2 + \xi^2} + y+b}{\sqrt{(y-b)^2 + \xi^2} + y-b}$$

The same procedure for the *Tyz* field yields

$$T_{yz} = G\sigma \left( \ln \frac{\sqrt{(x+a)^2+(y-b)^2+(z+H)^2}+x+a}{\sqrt{(x-a)^2+(y-b)^2+(z+H)^2}+x-a} \right.$$
$$+ \ln \frac{\sqrt{(x+a)^2+(y+b)^2+(z+H+d)^2}+x+a}{\sqrt{(x-a)^2+(y+b)^2+(z+H+d)^2}+x-a}$$
$$- \ln \frac{\sqrt{(x+a)^2+(y+b)^2+(z+H)^2}+x+a}{\sqrt{(x-a)^2+(y+b)^2+(z+H)^2}+x-a}$$
$$\left. - \ln \frac{\sqrt{(x+a)^2+(y-b)^2+(z+H+d)^2}+x+a}{\sqrt{(x-a)^2+(y-b)^2+(z+H+d)^2}+x-a} \right) \quad (4A)$$

An arbitrary horizontal position of the block with respect to the reference frame chosen can be taken into account in Eq.3A and Eq.4A by the following coordinate transform

$x \to x-X_0, \quad y \to y-Y_0$

where $X_0$ and $Y_0$ are the lateral coordinates of the centre of mass of the block. A different approach in obtaining *Txz* and *Tyz* gravity gradient components of an arbitrary prism can be found in (Dransfield, 1994).

**Appendix B: *Txz* and *Tyz* gravity gradients above a round disk**

By the use of the results obtained by Lockerbie *et al.* (1993) for an arbitrary hollow cylinder, and some tabulated integrals (Prudnikov *et al.*, 1986), one can show that the *Txz* and *Tyz* gravity gradient components above a round disk of the mass density $\sigma$, radius *a* and thickness *d* can be represented as follows

$$T_{xz} = G\sigma \frac{x-X_0}{r^2} \left\{ \sqrt{(H+z)^2+(a+r)^2}\left((2-A)K(A)-2E(A)\right) \right.$$
$$\left. - \sqrt{(H+d+z)^2+(a+r)^2}\left((2-B)K(B)-2E(B)\right) \right\} \quad (1B)$$

$$T_{yz} = G\sigma \frac{y-Y_0}{r^2} \left\{ \sqrt{(H+z)^2+(a+r)^2}\left((2-A)K(A)-2E(A)\right) \right.$$
$$\left. - \sqrt{(H+d+z)^2+(a+r)^2}\left((2-B)K(B)-2E(B)\right) \right\} \quad (2B)$$

where *G* is the gravitational constant, *H* is the disk top depth, $E(k)$ and $K(k)$ are the complete elliptic integrals of the first and the second kinds respectively, $X_0$ and $Y_0$ are the lateral coordinates of the centre of mass of the disk with respect to the reference frame chosen, and the following notations are used

$$A = \frac{4ar}{(H+z)^2 + (a+r)^2} \;,\; B = \frac{4ar}{(H+d+z)^2 + (a+r)^2} \;,\; r = \sqrt{(x-X_0)^2 + (y-Y_0)^2}$$